# Structural studies of silicate glasses with PbS and PbSe nanoparticles

Gurin V.S.
*Research Institute for Physical Chemical Problems, Belarusian State University, Minsk, Belarus*

Rachkovskaya G.E., Zakharevich G.B.
*Belarusian State Technological University*

Kichanov S.E.
*Joint Institute of Nuclear Research, Dubna, Russia*

**Abstract**

Optical materials with semiconductor nanoparticles within dielectric matrices are of interest for construction of non-linear optical elements, selective filters, spectral converters, etc. In the present work, we concern the glasses with PbS and PbSe fabricated by the two-step technique on the basis of silicate glass matrix and report the recent studied using the small-angle neutron scattering (SANS) technique. A proper understanding the structure of material, size of particles and spectral features need to control optical functionality of glasses. SANS is efficient non-destructive technique allowing a wide range of structural models in the size range from atomic clusters to submicron size. Here we combine SANS measurements with wide-angle X-ray diffraction (WXRD) and electron microscopy to get information on nanoparticles formed in a series of PbS and PbSe-doped silicate glasses.

## 1. Introduction

Glasses with quantum sized semiconductor nanoparticles are novel class of optical materials with advanced features for development of non-linear optical elements, passive laser locks, selective spectral filters, etc. [1,2] Lead chalcogenides (PbS and PbSe) are featured an explicit manifestation of quantum size effects even for ten-nm-scale nanoparticles due to large value of the Bohr exciton radius. The nanoparticles can be successfully localized within a glass matrix (various type: silicate, phosphate, borate), and for the size range 2-10 nm their excitonic absorption peaks are located in the near IR that is of great interest for application in lasers for telecommunication and medicine [1]. Meanwhile, the nanoparticles synthesized through colloidal methods with ligand stabilization acquire usually the less size range and their absorption covers the visible spectral range up to UV. In contrast to colloidal chalcogenide nanoparticles, the particles in glass matrices are stable with respect to environment as well to intense optical radiation that provide many possibilities to use glasses as non-linear optical media with proper control of spectral properties via the variation of lead chalcogenide particle state (size, concentration, aggregation, etc.). However, detailed structural studies of nanoparticles in glass are troubled because low particle concentration. Any extraction of particles from a glass matrix is impossible without their essential modification. Thus, indemolition techniques are of great interest to get new information on the state of nanoparticles in such systems. Small-angle scattering, in particular, neutron scattering, i.e. SANS, is one of such techniques that can provide versatile structural information on composite nanosystems at different scales. In the present paper, we combine SANS with WXRD and transmission electron microscopy (TEM) for a series of silicate glasses with PbS and PbSe nanoparticles.

## 2. Preparation of glasses

The samples of PbS/PbSe glasses have been fabricated according to the technique described by us earlier [4]. It is based on conventional two-step method used for semiconductor particles in glass: (1) a melting-cooling cycle of the glass-forming oxides ($SiO_2$, $Na_2O$, $K_2O$, $ZnO$, $Al_2O_3$, and PbO) with fluoride component doping followed by (2) secondary heat treatment near $T_g$ of the glass (480-530°C). Nanoparticles are nucleated and growing during the step (2), and temperature and duration of this step appears to be usual tool to control final state of the nanoparticles in the matrix.

# 3. Experimental results
## 3.1. WXRD

WXRD was used here to discover particles as nanocrystalline phases while the glass appears as typical amorphous constituent in this system and generates wide haloes and noises. Fig. 1-2 present the diffraction patterns for glasses with PbS and PbSe, respectively, JCPDS 05-0592 and JCPDS 06-0354. Small size of nanoparticles and overall low concentration do not allow good diffraction peaks, but the main positions may be detected for identification with cubic phases of PbS and PbSe. Approximate particle sizes can be also determined through the conventional Scherrer formula. The values for PbS are shown in Fig. 1, but for PbSe only order of magnitude may be noted due to strong noise and broadening, < 10nm.

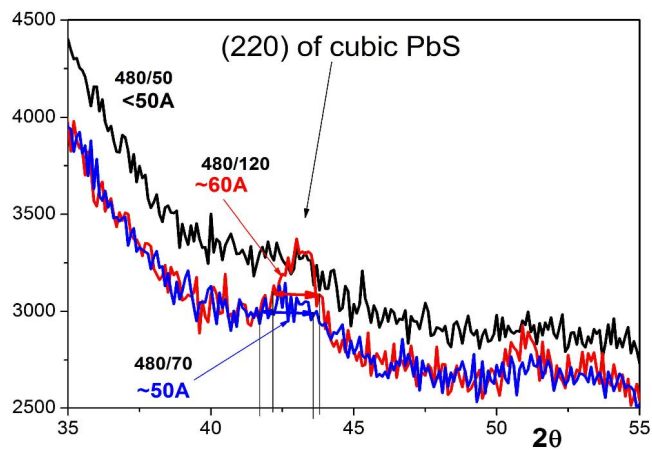

Fig. 1. Diffraction patterns for the glasses with PbS

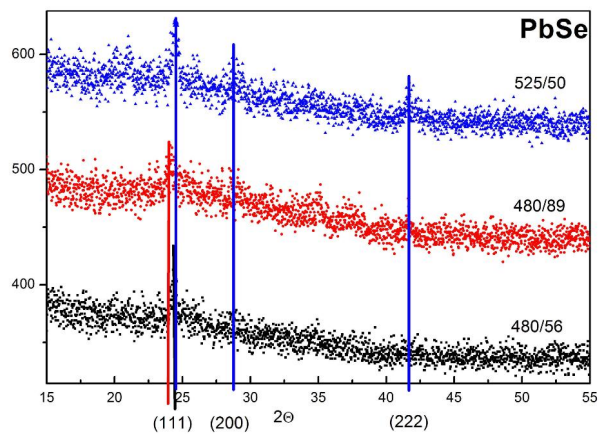

Fig. 2. Diffraction patterns for the glasses with PbSe

Table 1. Unit cell parameters of PbS and PbSe nanoparticles determined from WXRD measurements for a series of glasses after different secondary heat treatment

| Heat treatment, temperature/duration, °C/h | a, Å | $(a-a_{ref})/a_{ref}$, % | Particle size, nm |
|---|---|---|---|
| PbS - $a_{ref}$ = 5.9315 Å | | | |
| 480/49 | 5.936 | +0.08 | 5-10 |
| 480/56 | 5.943 | +0.19 | |
| 480/75 | ~5.95* | +0.3 | |
| 480/81 | 5.943 | +0.19 | |
| 480/105 | 5.935 | +0.06 | |
| 480/132 | 5.946 | +0.24 | |
| 480/150 | 5.935 | +0.06 | |
| PbSe - $a_{ref}$ = 6.1213 Å | | | |
| 480/56 | 6.3 | +2.9 | 8-10 |
| 480/89 | 6.131 | +0.16 | |
| 525/50 | 6.156 | +0.57 | |

Unit cell parameters of nanocrystals can be derived in these glasses in some cases when WXRD appear to be better. Table 1 collect these data for a series of PbS/PbSe – containing glasses. The values of **a** are rather close to the reference bulk counterparts, and a slight expansion can be noted, that, however, does not correlate with heat treatment duration. The expansion may be associated with glass-particle interaction, while for bare semiconductor nanoparticles a lattice contraction is more expectable.

*3.2. SANS*

Fig. 3,4 present the raw data of SANS and the processing of the scattering intensities by replotting them as $\log_{10}(I)$-$Q^2$ dependence. For the range of Q when QR<<1 (R is particle radius) it is admissible to use the Guinier approximation for to get the size of particles in the model of uniform isolated spheres.

$$I = (4\pi^2 N^2/3) \exp(-Q^2 R_g^2/3) \quad (1)$$

N is particle concentration and radius of gyration $R_g$ is related to the particle radius in this model as $R=(5/3)^{1/2}R_g$. However, the complicated profile of the SANS curves in the full interval suggests that this is only simplest basic model, and more elaborated geometry will be treated in future.

These data show that for the glasses both with PbS and PbSe the particle sizes depend on the variation of thermal treatment. For PbS-doped glasses they enter the interval of 3-6 nm, but for the glasses with PbSe the two fraction of particles may be evaluated (Table 2), and the heat treatment also provide the increase of particle radii. The concave shape of the full curves for some samples can argue on particle aggregation or formation of complex particle-matrix structures. However, there is no possiblility to conclude more within the framework of the Guinier approximation, and more detailed analysis should be done to simulate SANS for the whole system [3,6].

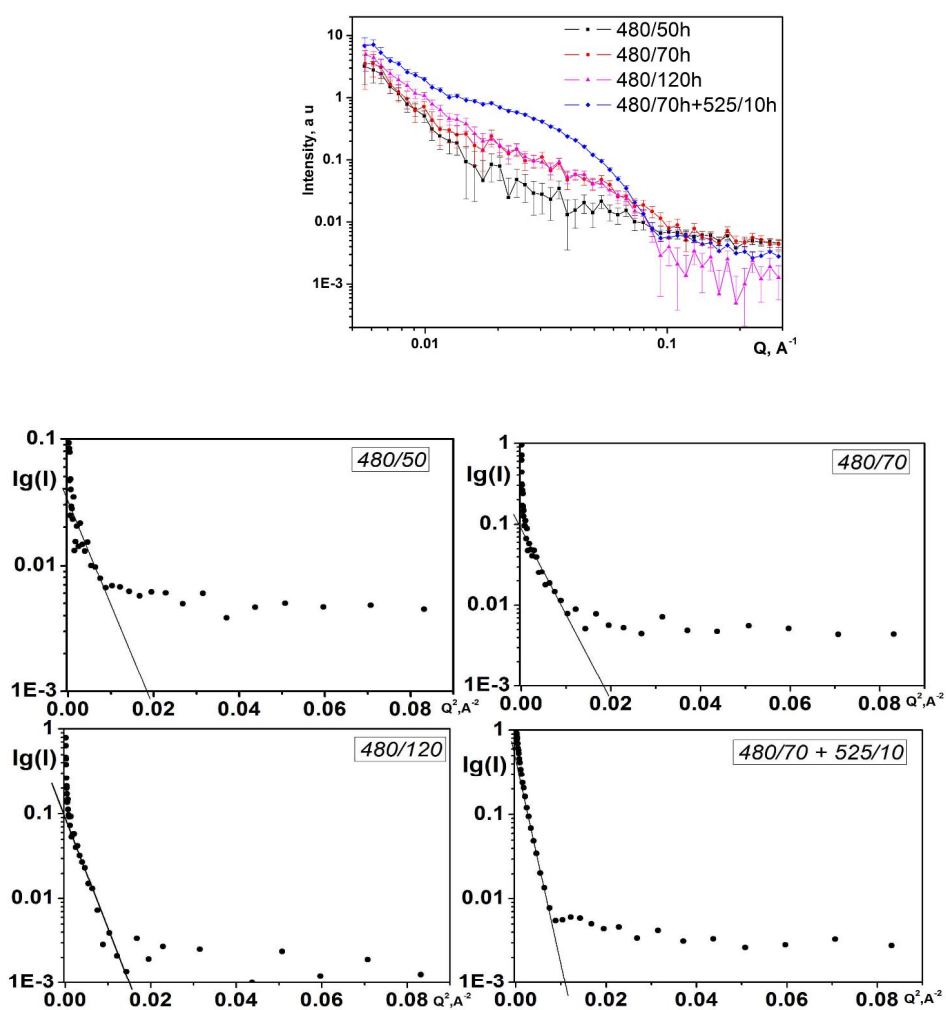

Fig. 3. SANS data for the glasses with PbS nanoparticles (top) and their approximation as $\log_{10}(I)$-$Q^2$ plots for derivation of particle size estimation through the Guinier formula (Table 2). Parameters of heat treatment of the glasses are labeled as 'temperature, °C/duration, h'.

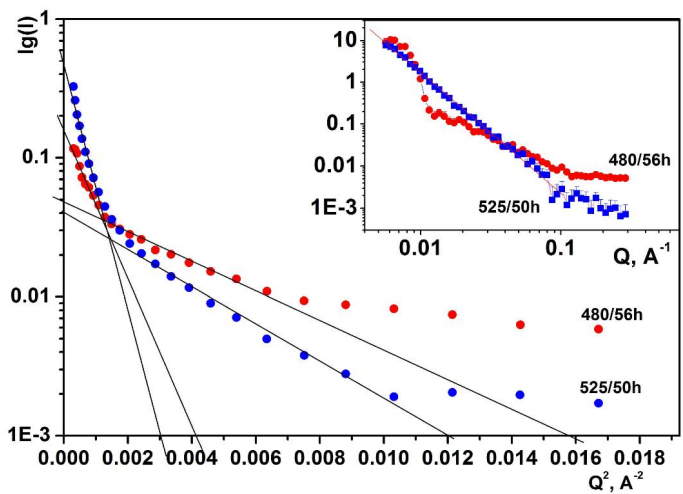

Fig. 4. SANS data for the glasses with PbSe nanoparticles (inset) and their approximation as $\log_{10}(I)$-$Q^2$ plots for derivation of particle size estimation through the Guinier formula (Table 2). Parameters of heat treatment of the glasses are labeled as 'temperature, °C/duration, h'.

Table 2. Values of particle sizes evaluated through the Guinier approximation for PbS- and PbSe-doped glasses. Parameters of heat treatment of the glasses are labeled as 'temperature, °C/duration, h'.

| Glass sample | $R_g$, Å | R, Å | Approx. diameter, nm |
|---|---|---|---|
| PbS | | | |
| 480/50 | 23.5 | 30.3 | 6 |
| 480/70 | 27.8 | 35.9 | 7 |
| 480/120 | 30.5 | 39.3 | 8 |
| 480/70+525/10 | 42.5 | 54.8 | 11 |
| | | | |
| PbSe | | | |
| 480/56 | 27.1 | 35.0 | 7 |
| | 62.4 | 80.5 | 16 |
| | | | |
| 525/50 | 31.3 | 40.4 | 8 |
| | 78.4 | 101.1 | 20 |

## 3.3. TEM

TEM data (Fig. 5) within the framework of this presentation are displayed by typical views both for PbS- and PbSe- containing glasses. Occurrence of nanoparticles at the clear contrast within the background of glass evidences formation of new solid nanophases rather than any homogeneous glass in consistence with WXRD results, although the latter were shown to appear noisy and broadened. Numerically, the size range of nanoparticles in TEM views and WXRD are close, ten nanometers and less. Partial aggregation of nanoparticles can explain some part of the larger particles in the micrographs. A study of sequence of TEM data for the glasses with various temperature and time of the secondary heat treatment is expected to reveal possible laws for their thermostimulated growth.

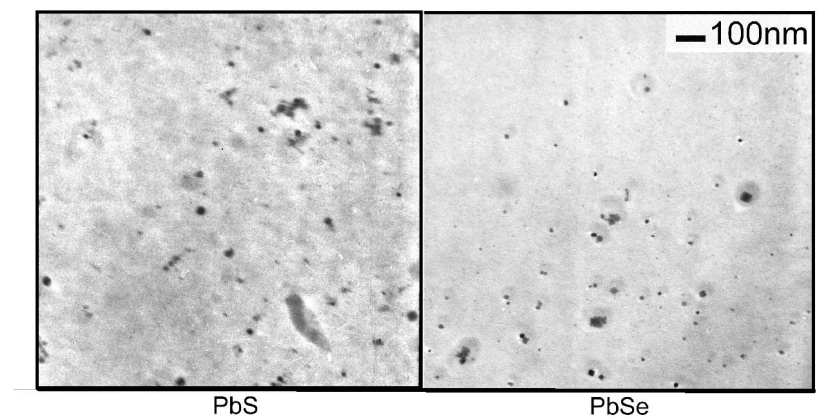

Fig. 5. Typical micrographs of PbS and PbSe nanoparticles in the glasses under study.

As for SANS, we also can notice the general consistence of the size ranges, and additively, more complicated version of size distribution for the glasses with PbSe, may be revealed in the TEM micrograph. On the other hand, the values of size of these nanoparticle derived from the optical data on excitonic absorption and photoluminescence are also in good correspondence [4,5].

## 4. Conclusion

Silicate glasses with PbS and PbSe nanoparticles have been studied with WXRD, SANS, and TEM. The results of each technique provide structural features of particles and glass matrix. The nanoparticles of the diameter range of 5-12 nm are detected indicating general consistence of the different measurement. The nanoparticles are quantum dots in the strong quantization mode. Their partial aggregation appears as the second level of ordering in these materials.